\documentclass[twocolumn, pra, showpacs]{revtex4}
\usepackage{graphicx}
\usepackage{dcolumn}
\usepackage{bm}
\usepackage{amsmath}

\begin{document}

\title{Dynamics of Two-Component Bose-Einstein Condensates Coupled with Environment}
\author{Yajiang Hao}
\email{haoyj@ustb.edu.cn}
\address{Department of Physics, University of Science and Technology
Beijing, Beijing 100083, P. R. China}
\author{Qiang Gu}
\email{qgu@ustb.edu.cn}
\address{Department of Physics, University of Science and Technology
Beijing, Beijing 100083, P. R. China}
\date{\today }

\begin{abstract}
We investigate the dynamics of an open Bose-Einstein condensate
system consisting of two hyperfine states of the same atomic species
which are coupled by tunable Raman laser. It is already suggested
that the detuning between the laser frequency and transition
frequency affect significantly on the dynamics of the pure
condensate. Here we show that the detuning effect is suppressed by
noise and dissipation caused by the environment. The increase of
coherence and purity are also displayed for specific parameters. As
a verification to the lowest-order approximation we derive the
hierarchy of motion equations in the second-order approximation. It
turns out that the former one can describe the dynamical evolution
qualitatively for weak noise and dissipation and quantitatively for
strong noise and dissipation.
\end{abstract}
\maketitle


\preprint{APS/123-QED}

\narrowtext

\section{introduction}
With the rapid experimental progress in manipulation of quantum
gases multi-component atomic gases have now been an active research
field in cold atomic physics. Since Myatt et al. first produced a
binary mixture of condensates consisting of two hyperfine states
($|F=1,m_f=-1\rangle$ and $|F=2,m_f=2\rangle$) of $^{87}$Rb
\cite{Myatt}, the two-component atomic bosons of $^{87}$Rb in
$|F=1,m_f=-1\rangle$ and $|F=2,m_f=1\rangle$ states received
intensive study in experiments subsequently
\cite{Matthews98,Hall98a,Hall98b,Lewan02,Erhard,Zibold}. A variety
of dynamical behaviors have been observed both in the Bose-Einstein
condensate (BEC) \cite{Matthews98,Hall98a,Hall98b} and in the
noncondensed sample \cite{Lewan02}. Meanwhile, theoretical works
have predicted that the two-component Bose gas may exhibit exotic
ground state and vortex structures \cite{Ho96,Pu,Ao,Cazalilla,Zhou}
as well as interesting dynamical properties, such as the Rabi
oscillation of population between the two states\cite {Matthews98},
nonlinear population dynamics \cite{Will99} and quantum
self-trapping \cite{ZDChen}.

In the two-component atomic gas, the two hyperfine states
($|F=1,m_f=-1\rangle$ and $|F=2,m_f=1\rangle$) can be coupled by a
two-photon transition which converts the $^{87}$Rb atom from one
state to the other \cite{Matthews98,Hall98a,Hall98b,Zibold}. This
transition thus drives the population dynamics. In this case, the
two-component system is analogous to the Bose gas trapped in the
double-well and both of them have also been investigated intensively
in recent years \cite{ZDChen,DWTheory,DWExp}. The dynamical
properties of the condensate satisfy the time-dependent
Gross-Pitaevskii (GP) theory because the Bose atoms condensed in the
ground state and the system can be described by the macroscopic
wavefunction.

However, under experimental conditions, the condensates should be
regarded as an open system coupled with the environment since the
condensate usually coexists with noncondensed thermal cloud and atom
loss is also unavoidable. In this situation, noise and dissipation
may play important roles. Many theoretical works have discussed the
noise and dissipation-induced effects in the open double-well
condensates recently \cite{DFWalls,Pitaevskii,Boukobza,AVardi}, and
indicated that negligible changes in dynamical properties of the
condensate appear, such as the decay of quantum self-trapping
\cite{DFWalls}, decoherence \cite{Pitaevskii} and dephasing
\cite{Boukobza,AVardi}. The dephasing phenomenon has even been
observed in experiments \cite{RGati}. However, sometimes
dissipations may result in enhancement of the quantum effect. For
example, a thermally enhanced quantum-oscillation was observed
during the domain formation process in the ferromagnetic spin-1
condensate \cite{Gu}. Li et al. also found that the maximum spin
squeezing can be reached even in presence of particle losses
\cite{YLi}. Most recently, Witthaut et al. showed that the
dissipation could lead to enhancement of coherence in the
double-well BECs under specific conditions \cite{Witthaut}.

In this paper we will investigates the open two-component BECs.
Although this system can be mapped into a double-well condensate
model, it is of interest in its own right. For example, dynamics of
this system can be controlled by tuning the laser frequency by which
the two internal states of atoms are coupled \cite{Will99,ZDChen}.
So our main purpose is to discuss the noise and dissipation-induced
effects when the two components are coupled by the detuned laser.
The Hamiltonian is expressed using the pseudo-angular-momentum
operators and the dynamical properties is evaluated by a hierarchy
of ordinary differential equations of Bloch vectors, which are
formulated on the basis of master equation method \cite{Witthaut}.

The paper is organized as follows. Section II formulates equations
of motion in the first-order approximation for an open two-component
Bose condensate with noise and dissipation. Section III discusses
dynamics of Bloch vectors for both the closed and open systems,
respectively. Section IV is devoted to evolutions of the coherence
and purity for certain given parameters. In Section V, we extend the
motion equations to the second-order approximation and compare the
results with those in the first order approximation. A summary is
given in the last section.

\section{model and method}

The two-component BEC is described by the Hamiltonian of second
quantization in the single mode approximation (natural unit $\hbar
=1$ is used throughout the paper)
\begin{equation}
\hat{H}=\sum_{j=1,2}\hat{H}_j+\hat{H}_{int}+\hat{H}_f,
\end{equation}
where
\[
\hat{H}_j=\omega _j\hat{n}_j+\frac{g_j}2\hat{n}_j^2
\]
with $\omega _j=\int d^3r\phi _j^{*}\left( r\right) \left[ -\frac{\hbar ^2}{%
2m}\nabla ^2+V_j\left( r\right) \right] \phi _j\left( r\right)$. The
interaction term between the two hyperfine states of atoms takes the
form of
\[
\hat{H}_{int}=g_{12}\hat{n}_1\hat{n}_2,
\]
and the transition term is
\[
\hat{H}_f=G\left( \hat{a}_1^{\dagger }\hat{a}_2e^{i\varphi \left( t\right) }+%
\hat{a}_2^{\dagger }\hat{a}_1e^{-i\varphi \left( t\right) }\right) .
\]
Here $g_j$ ($j$=1,2) and $g_{12}$ are the effective intra- and
inter-components interaction constants that can be controlled
experimentally by tuning the atomic $s$-wave scattering length with
Feshbach resonance techniques. The transition probability between
both components with a small detuning is denoted by the Josephson
coupling strength $G$. In the rotating wave approximation the phase
$\varphi \left( t\right) =\Delta t$, where $\Delta $ is the detuning
between coupling laser frequencies and transition frequency
$|\omega_1-\omega_2|$ between two components.

The Hamiltonian can be reexpressed by introducing the Schwinger
pseudo-angular-momentum operators defined as
$\hat{L}_x=\frac 12\left( \hat{a}_1^{\dagger }\hat{a}_2+ \hat{a}_2^{\dagger }%
\hat{a}_1\right)$, $\hat{L}_y=\frac 1{2i}\left( \hat{a} _1^{\dagger }\hat{a}%
_2-\hat{a}_2^{\dagger }\hat{a}_1\right)$, $\hat{L} _z=\frac 12\left( \hat{a}%
_1^{\dagger }\hat{a}_1-\hat{a}_2^{\dagger }\hat{a}_2\right)$ with
Casimir
invariant $\hat{L}^2=\frac{\hat{N}}2\left( \frac{\hat{N}}2+1\right)$, where $%
\hat{N}=\hat{a}_1^{\dagger }\hat{a}_1+ \hat{a}_2^{\dagger
}\hat{a}_2$ is the total boson number operator. $\hat{L}_{\pm
}=\hat{L}_x\pm i\hat{L}_y$ and $\hat{L}_z$ satisfy the usual angular
momentum commutation relation: $\left[ \hat{L}_z,\hat{L}_{\pm
}\right] =\pm \hat{L}_{\pm },$ and $\left[ \hat{L}_{+},\hat{L}_{-}\right] =2%
\hat{L}_z$. Therefore the Hamiltonian can be written as
\begin{equation}
\hat{H}=\omega _0\hat{L}_z+q\hat{L}_z^2+G\left(
\hat{L}_{+}e^{i\varphi \left( t\right) }+\hat{L}_{-}e^{-i\varphi
\left( t\right) }\right) \label{Hamil}
\end{equation}
with $\omega _0=\omega _1-\omega _2+\left( N-1\right) \left(
g_1-g_2\right) /2$ and $q=\left( g_1+g_2\right) /2-g_{12}$. In the
Bose gas of $\left|F=1,m_f=-1\right\rangle$ and
$\left|F=2,m_f=1\right \rangle$ states of $^{87}$Rb, the effective
interaction constants are known to be in the proportion
$g_1:g_{12}:g_2=1.03:1:0.97$ \cite{Matthews98}. These parameters can
be tuned by magnetic Feshbach resonance \cite{SBPapp} so that they
can  be more different.

With the Schwinger representation the general solution of the
Schr\"odinger equation $i\hbar
\frac{\partial{|\psi(t)\rangle}}{\partial t}=\hat{H}|\psi(t)\rangle$
governing the dynamics of system can be obtained with a
time-dependent unitary transformation \cite{ZDChen} and when the
system couples with the environment, its dynamics will reserve to
the master equation
\begin{eqnarray}\label{Rho}
\dot{\hat{\rho}} &=&-i\left[
\hat{H},\hat{\rho}\right]+\mathcal{L}_p+ \mathcal{L}_a,
\end{eqnarray}
where $\hat{\rho}$ is the density operator,
$\mathcal{L}_p=-\frac{\gamma _p} 2\sum_{j=1,2}\left( n_j^2\rho +\rho
n_j^2-2n_j\rho n_j\right)$ denotes the noise term and
$\mathcal{L}_a=-\frac 12\sum_{j=1,2}\gamma _{a_j}\left( \hat{a}
_j^{\dagger }\hat{a}_j\rho +\rho \hat{a}_j^{\dagger
}\hat{a}_j-2\hat{a} _j\rho \hat{a}_j^{\dagger }\right)$ is the
dissipation term. Accordingly, $\gamma_p$ and $\gamma_{a_j}$
describe, respectively, the strength of noise and the dissipation
velocity for the component $j$. The one-time average of a system
operator $\hat{A}$ can be calculated by $\left\langle \hat{A}
\right\rangle=\text{tr}\left[ \hat{A}\hat{\rho}\left( t\right)
\right]$. Here we define the single-particle Bloch vector and
particle number as $s_j\left( t\right) =2\text{tr}\left[
\hat{L}_j\hat{\rho}\left( t\right) \right]$ and $ n\left( t\right)
=\text{tr}\left[ \left( \hat{n}_1+\hat{n}_2\right) \hat{ \rho}\left(
t\right) \right]$, while the time derivatives of them are defined as
$\dot{s}_j\left( t\right) =2\text{tr}\left[\hat{L}_j\dot{\hat{\rho}}
\left( t\right) \right]$ and $\dot{n}\left( t\right)
=\text{tr}\left[n\dot{ \hat{\rho}}\left( t\right) \right]$.

Insert the Hamiltonian $\hat{H}$ into Eq. (\ref{Rho}), we will find
that the first-order operators $\hat{L}_j$ depend not only on
themselves, but also on the second-order moments $\left\langle
\hat{L}_i\hat{L}_j\right\rangle$. Similarly, the time evolution of
the second-order moments depends on third-order moments, and so on.
In order to obtain a closed set of equations of motion, the
hierarchy of equations must be truncated at some stage by
approximating the $N$th order expectation value in terms of all
lower-order moments. Here we take the first-order approximations, i.e., $%
\left\langle \hat{L}_i\hat{L}_j\right\rangle \approx \left\langle
\hat{L} _i\right\rangle \left\langle \hat{L}_j\right\rangle$, and
thus the equations of motion for the Bloch vector take the
formulation of
\begin{eqnarray}
\dot{s}_x &=&-\omega _0s_y-qs_ys_z-2Gs_z\sin \varphi \left( t\right)
-T_2^{-1}s_x,  \nonumber \\
\dot{s}_y &=&\omega _0s_x+qs_xs_z-2Gs_z\cos \varphi \left( t\right)
-T_2^{-1}s_y,  \nonumber \\
\dot{s}_z &=&2G\left( s_x\sin \varphi \left( t\right) +s_y\cos
\varphi
\left( t\right) \right) -T_1^{-1}s_z-T_1^{-1}f_an,  \nonumber \\
\dot{n} &=&-T_1^{-1}n-T_1^{-1}f_as_z.  \label{1stDE}
\end{eqnarray}
Here the damping parameters $T_1$, $T_2$ and the relative
dissipation velocity $f_a$ are expressed as
\begin{eqnarray*}
T_1^{-1} =\frac 12\left( \gamma _{a_1}+\gamma _{a_2}\right),
T_2^{-1} =\gamma _p+T_1^{-1}, f_a=\frac{\gamma _{a_2}-\gamma
_{a_1}}{\gamma _{a_2}+\gamma _{a_1}}.
\end{eqnarray*}
The dissipation can be controlled artificially by shining a laser
beam onto the condensates or be achieved by a forced radio frequency
transition to an untrapped magnetic substate \cite{IBloch}. The
relative dissipation rate $f_a$ can be changed from 0 to 1 and we
take a medium value 0.5 in the present paper if it is not given
particularly when there exists atom loss.

In this work we will investigate the phase coherence between two
species and purity of the system, both of which are related with the
Bloch vector by $ \alpha \left( t\right) =2\left| \left\langle
\hat{a}_1^{\dagger } \hat{a} _2\right\rangle \right|/\left\langle
\hat{n}_1+\hat{n} _2\right\rangle=\sqrt{ s_x ^2+s_y ^2}/ n$ and
$p\left(t\right)=|\mathbf{s}|^2/n^2$ such that the dynamics of them
can be investigated by solving Eq. (\ref{1stDE}) for the definite
initial condition. For convenience in the following evaluation the
initial conditions $\mathbf{s}(0)=(s_x(0),s_y(0),s_z(0))$ will be
taken as $N(a,0,\sqrt {1-a^2})$ with $N$ being the initial particle
number in the system. $s_x(0)=0$ and $s_z(0)=N$ ($a$=0) corresponds
to the case that all atoms occupy in the same hyperfine state while
$s_x(0)=N$ and $s_z(0)=0$ ($a$=1) corresponds to the case that atoms
populate equally in two hyperfine states. In the following
evaluation we will consider the system of $N=1000$.

\section{Dynamics of Bloch vector}

\begin{figure}[tbp]
\includegraphics[width=0.4\textwidth]{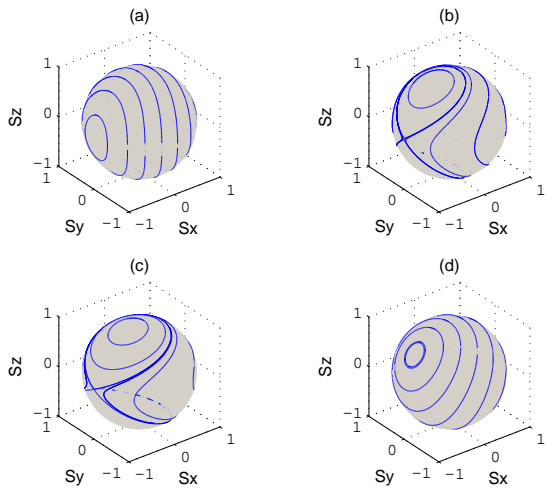}\newline
\caption{The evolving Bloch vector for the closed system of $N=1000$
and $\Delta=0$. (a) $q$=0.001, $\omega_0$=0.0; (b) $q$=0.005, $
\omega_0$=0.0; (c) $q$=0.008, $\omega_0$=0.0; (d) $q$=0.001,
$\omega_0$=1.0.} \label{fig1}
\end{figure}

\begin{figure}[tbp]
\includegraphics[width=0.4\textwidth]{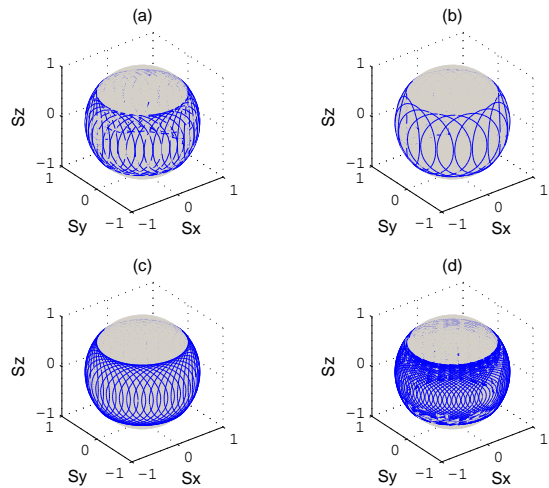}\newline
\caption{The evolving Bloch vector for the closed system of $N=1000$
and $\omega_0$=0.0. The initial state is chosen as $s_x(0)=0.6N$.
(a) $q$=0.0, $\Delta$ =0.05; (b) $q$=0.0, $\Delta$=0.1; (c)
$q$=0.001, $\Delta$=0.05; (d) $q$ =0.01, $\Delta$=0.05.}
\label{fig2}
\end{figure}
In the present Schwinger representation all interesting physical
quantities can be formulated with Bloch vector. For example, for
particle number conserved system coherence is the module of the
projection of Bloch vector on the $s_x$-$s_y$ plane, its purity is
the module square of Bloch vector and $s_z(t)$ corresponds to the
population imbalance between two components. So the dynamical
evolution of Bloch vector is worth to study very much. This section
will focus on the evolving Bloch vector and the explicit
investigation of coherence and purity will be given in the next
section.

\subsection{The Closed System}

Firstly, we investigate the dynamics of the system without coupling
with environment, i.e., there is not noise and particle dissipation.
In Fig. 1 (a-c) we display the Bloch vector for different nonlinear
constants $q$ for the case of detuning $\Delta=0$. The Bloch vector
evolves periodically in the surface of Bloch sphere and forms a
closed orbital, while different initial condition corresponds to
respective orbital.

For the weak nonlinear constants the orbitals behave as an almost
perfect circular orbital parallel to the plane of $s_x=0$ and the
fixed point belong to the line ($s_y=s_z=0$). That is to say, atoms
shall oscillate periodically between these two hyperfine states and
in each period the average occupation probability is equal in both
states. With the increase of nonlinear term the original circle
orbital shall deform and their centers moved from the position of
$s_z=0$ ($q=0$) to the north pole of $s_z=1$. The fixed points move
to the upper sphere. For example, $s_z(t)$ approaches to 1.0 for all
time as $q=0.008$. This is referred to macroscopic quantum
self-trapping (MQST) that atoms will prefer to stay in the same
hyperfine state instead that during an oscillation period atoms
occupy in equal probability in both states. The transition between
periodical oscillation and MQST corresponds to the orbital similar
to lemniscates spreading in the surface of Bloch sphere.

It deserves to notice that in Ref. \cite{Zibold} those new formed
two trajectories and fixed points $F_{\pm}$ for the strong
interaction correspond to $\pm z_0$, respectively, while in the
present evaluation we only choose the positive $z_0$. In addition
the transition from the periodical oscillation to MQST can be
induced not only by the increased interaction but also the increased
initial imbalances between two states.

The effect of $\omega_0$ can be displayed by comparing Fig. 1a
($\omega_0$=0) and Fig. 1d ($\omega_0$=1.0). It is shown that its
value shall change the orientation of the orbital plane. The
increase of $\omega_0$ leads to the turn to the north pole of the
orientation of orbital plane, which is also an expression of MQST.
In a word, for the closed system of nondetuning MQST can exhibit by
tuning $\omega_0$, $q$ and initial imbalance, whose orbital center
deviate from $s_z=0$.

In Fig. 2 we display the orbitals for the situation where there
exist detune between the transition frequency and the coupling
laser. In this case the Bloch vector will no longer evolve
periodically and its path behave as a spiral line in the surface of
Bloch sphere. With the time delay the path of Bloch vector wind each
other and looks like a basket when the evolving time is long enough.
By comparing Fig. 2a with Fig. 2b and Fig. 2c with Fig. 2d we found
that the sparser the net grid, the bigger the detuning and the
denser the net grid, the stronger the nonlinear constant. The denser
track means that the evolution of Bloch vector and thus the dynamics
of two-component BECs is more close to the pure periodicity in short
time. In spite of the complicated orbital its projection on $s_z$
axis is still periodical oscillation while the period is larger a
little than the case without detune. The evaluation show that its
period become longer although the atoms still transit periodically
between two states and the transition point between delocalization
and self-trapping also deviate slightly from the case without
detune.

\begin{figure}[tbp]
\includegraphics[width=0.4\textwidth]{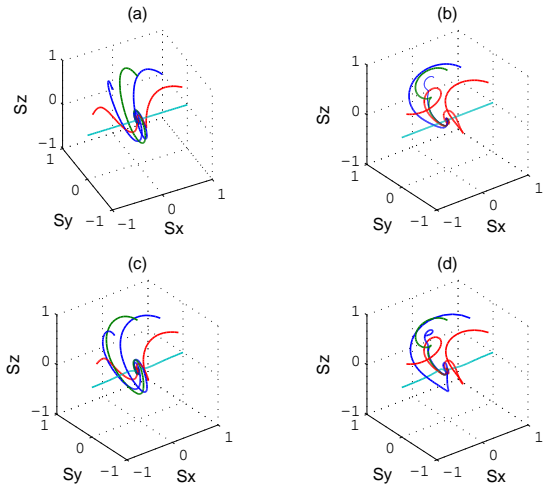}\newline
\caption{The evolving Bloch vector for the system of $N=1000$
subject to noise for $\omega_0$=0.0, $\gamma_p$=1.0 and
$T_1^{-1}$=0.0. (a) $q$ =0.00, $\Delta$=0.0; (b) $q$=0.005,
$\Delta$=0.0; (c) $q$=0.001, $\Delta$ =0.05; (d) $q$=0.005,
$\Delta$=0.05.} \label{fig3}
\end{figure}

\subsection{The System Coupling With Environment}

After investigating the dynamics of closed system, we focus on the
system coupling with environment. The effects of noise and
dissipation are shown in Fig. 3 and in Fig. 4, respectively. Here
different orbitals in Bloch space correspond to different initial
condition.

For the system subject to noise the module of Bloch vector will not
be conserved rather decrease to zero in a spiral during the
dynamical evolution. It would not be on the surface of Bloch sphere
rather shrink to the original point of coordinate frame even for
very weak noise, whose strength only affect the evolving time to the
original point. That is to say, when suffering noise instead of
dissipation the atom number populated in two components would always
tend to balance. In addition the purity of system tend to zero that
corresponds to the module of Bloch vector. This will be discussed
further in the later section. By comparing Fig. 3a with Fig. 3b and
Fig. 3c with Fig. 3d we found that the stronger nonlinear
interaction constant and large detuning shall induce that the Bloch
vector run as an irregular orbital.

In Fig. 4 we plot the orbitals for the case that there are only
dissipation instead of noise for different nonlinear interaction
constants $q$. The Bloch vector shall be always on the surface of
Bloch sphere but the orbitals become irregular and not closed. In
this situation for different initial atom population in two
components the final state shall be indefinite completely. With the
increase of dissipation the atomic loss rate become faster and the
lifetime of BECs become shorter and shorter. This means that along
with the BECs oscillation between two hyperfine states
aperiodically, more and more atoms escape from the system. As the
interaction constant increases the shape of orbitals become more and
more complicated.

\begin{figure}[tbp]
\includegraphics[width=0.4\textwidth]{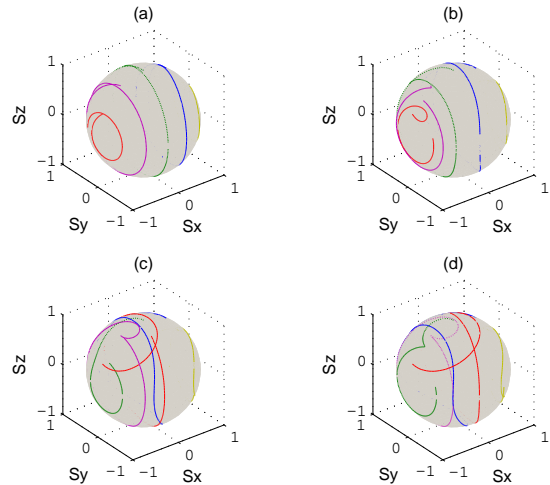}\newline
\caption{The evolving Bloch vector for the system of $N=1000$
subject to dissipation for $\omega_0$=0.0, $\gamma_p$=0.0
$\Delta$=0.05 and $T_1^{-1}$ =1.0. (a) $q$=0.001; (b) $q$=0.005; (c)
$q$=0.008; (d) $q$=0.01.} \label{fig4}
\end{figure}

\begin{figure}[tbp]
\includegraphics[width=0.4\textwidth]{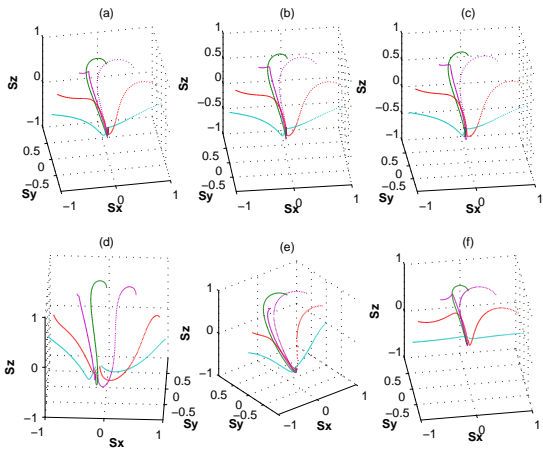}\newline
\caption{The evolving Bloch vector for the system of $N=1000$
subject to both noise and dissipation for $\omega_0$=0.0 and
$q$=0.005. (a)-(c) $\gamma_p$=2.0, $T_1^{-1}$=1.0; $\Delta$=0.0 (a),
0.05 (b), 0.1 (c). (d) $\gamma_p$=1.0, $T_1^{-1}$ =2.0,
$\Delta$=0.1. (e) $\gamma_p$=2.0, $T_1^{-1}$=2.0, $\Delta$=0.1. (f)
$\gamma_p$=2.0, $T_1^{-1}$=2.0, $\Delta$=0.1, $f_a$=0.0.}
\label{fig5}
\end{figure}

Fig. 5 displays the evolution of Bloch vector for the system of 1000
atoms and $q=0.005$ subject to both noise and dissipation. Here
$f_a$=0.5 in Fig. 5 (a)-(e) and $f_a$=0.0 in Fig. 5f. It is shown
that for different initial states it will always evolve to the same
point in the Bloch space, which correspond to the same final state,
while the change of noise and dissipation shall determine its
evolving path and dissipation velocity to the final state. In
addition the joint effect on the BECs induces that the evolving path
to the final state become more simple than the system subject to
only noise or dissipation. According to Fig. 5 (a)-(c)
($\Delta$=0.0, 0.05, 0.1) the detuning between transition frequency
and the frequency of laser induce no explicit effect. According to
Fig.5e and Fig. 5f, the relative dissipation rate of each components
shall result in different final position in Bloch space, i.e.,
different final states.

To sum up, noise always induce that atoms populate in two components
evenly and atom dissipation determine the lifetime of condensate. Both noise
and dissipation shall result in the irregular evolving path
of Bloch vector, i.e., the aperiodic dynamics. In this situation the
system always arrive at the same final state and the evolving path shall
be more simple compared with the case subject to only noise or dissipation.

\section{coherence and purity}

\begin{figure}[tbp]
\includegraphics[width=0.4\textwidth]{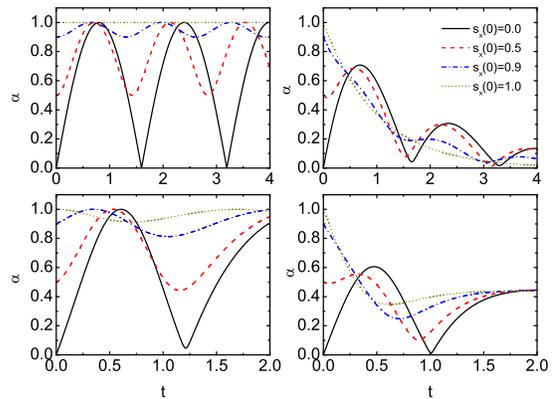}\newline
\caption{The coherence dynamics for the system of q=0.001,
$\omega_0$=0, $G_0$=1, $\Delta$=0 and $f_a$ =0.5. (a) $\gamma$=0.0,
$T_1^{-1}$=0.0; (b) $\gamma$=1.0, $T_1^{-1}$=0; (c) $\gamma$ =0.0,
$T_1^{-1}$=2.0; (d) $\gamma$=2.0, $T_1^{-1}$=3.0.} \label{fig6}
\end{figure}
\begin{figure}[tbp]
\includegraphics[width=0.4\textwidth]{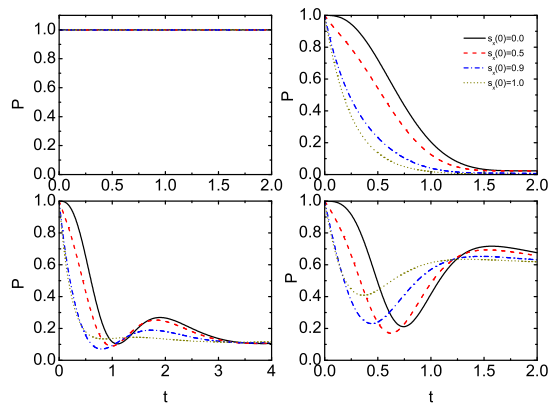}\newline
\caption{The time evolution of purity for the system of q=0.001,
$\omega_0$=0, $G_0$=1, $\Delta$=0 and $f_a$ =0.5. (a) $\gamma$=0.0,
$T_1^{-1}$=2.0; (b) $\gamma$=2.0, $T_1^{-1}$=0; (c) $\gamma$ =2.0,
$T_1^{-1}$=1.0; (d) $\gamma$=2.0, $T_1^{-1}$=3.0.} \label{fig7}
\end{figure}

In this section we will study the evolution of coherence and purity
for different noise and dissipation.

In Fig. 6 the evolution of coherence is displayed for different
initial conditions. It is shown that for a closed system the
coherence will always oscillate periodically as long as the initial
population difference in each component deviates from zero
($a\neq1$) while it will preserve the strongest coherence when the
atoms initially populate equally at two components ($a=1$) (Fig.
6a). When the BECs suffer noise instead of dissipation the coherence
will decrease oscillationly (Fig. 6b), and when there exist
dissipation instead of noise the coherence still oscillates
quasi-periodically (Fig. 6c). For the system exposed to both noise
and dissipation the coherence show rich evolving dynamics determined
by the initial populations (Fig. 6d). For the case of equal
population initially ($a$=1) the coherence between two components
decrease firstly and then increase, which is similar to the
stochastic resonance. For the polarized case initially ($a$=0) the
coherence show a process of first increase from zero to the
strongest value and then following a stochastic resonance. While for
the initial population between these two limits ($0<a<1$) coherence
will show the evolving process of first increase or decrease and
then following a stochastic resonance.

In short, particle loss do not temper with the coherence dynamics greatly
and the obvious effect of noise is the reduction of the maximum coherence.
So by tuning the strength of noise and dissipation rate we always can
observe the process of coherence enhancement. The similar effect on the
purity can also be seen (Fig. 7), where dissipation show no effect on purity.

The evolving purity is shown in Fig. 7. According to Fig. 7a
($\gamma=0$) and Fig. 7b ($T_1^{-1}=0$) the dissipation shall not
bring about the change of purity as long as the noise term is equal
to zero and the noise shall result in the decrease of purity. By
tuning the relative strength of noise and dissipation rate the
purity increase again after rapid decrease. It turns out that for
suitable parameter the purity can be enhanced to a very large value.
For instance, it reach 0.7 for $\gamma$=2.0, $T_1^{-1}$=3.0. In
addition, the purity evolving do not show qualitative difference for
different initial condition. It deserves to notice that the relative
dissipation rate $f_a$ play a critical role on whether the
enhancement of coherence and purity can be displayed. Generally, the
bigger $f_a$ shall result in the larger coherence and purity so we
take a medium value of $f_a=0.5$ here.

\section{second-order approximation}

In order to check the accuracy of the lowest-order approximation, we
now take into account the second-order correction and compare the
results from both cases. Up to the second-order approximation, the
hierarchy equations are truncated by approximating the third-order
expectation value $<\hat{L}_i\hat{L }_j\hat{L}_k>$ as the following:
\begin{eqnarray*}
\left\langle \hat{L}_i\hat{L}_j\hat{L}_k\right\rangle  &=&\left\langle \hat{L%
}_i\hat{L}_j\right\rangle \left\langle \hat{L}_k\right\rangle
+\left\langle
\hat{L}_i\right\rangle \left\langle \hat{L}_j\hat{L}_k\right\rangle  \\
&&+\left\langle \hat{L}_i\hat{L}_k\right\rangle \left\langle \hat{L}%
_j\right\rangle -2\left\langle \hat{L}_i\right\rangle \left\langle \hat{L}%
_j\right\rangle \left\langle \hat{L}_k\right\rangle .
\end{eqnarray*}
Thus the hierarchy of motion equation shall be composed of the Bloch
vector and the second-order moments $\Delta _{ij}=
4(<\hat{L}_i\hat{L}_j+\hat{L}_j\hat{L}
_i>-2<\hat{L}_i><\hat{L}_j>)$. They can be formulated as below
\begin{eqnarray*}
\dot{s}_x &=&-\omega _0s_y-\frac q2\left( \Delta
_{yz}+2s_ys_z\right)
-2Gs_z\sin \varphi \left( t\right) -T_2^{-1}s_x, \\
\dot{s}_y &=&\omega _0s_x+\frac q2\left( \Delta _{xz}+2s_xs_z\right)
-2Gs_z\cos \varphi \left( t\right) -T_2^{-1}s_y, \\
\dot{s}_z &=&2G\left( s_x\sin \varphi \left( t\right) +s_y\cos
\varphi
\left( t\right) \right) -T_1^{-1}s_z-T_1^{-1}f_an, \\
\dot{n} &=&-T_1^{-1}n-T_1^{-1}f_as_z, \\
\dot{\Delta}_{xx} &=&-2\omega _0\Delta _{xy}-4G\sin \varphi \left(
t\right)
\Delta _{xz}-2q\left[ \Delta _{xz}s_y+\Delta _{xy}s_z\right]  \\
&&-2T_2^{-1}\Delta _{xx}+2\left( T_2^{-1}-T_1^{-1}\right) \left(
\Delta
_{yy}+2s_y^2\right) +2T_1^{-1}n, \\
\dot{\Delta}_{yy} &=&2\omega _0\Delta _{xy}-4G\cos \varphi \left(
t\right)
\Delta _{yz}+2q\left( \Delta _{xy}s_z+\Delta _{yz}s_x\right)  \\
&&-2T_2^{-1}\Delta _{yy}+2\left( T_2^{-1}-T_1^{-1}\right) \left(
\Delta
_{xx}+2s_x^2\right) +2T_1^{-1}n, \\
\dot{\Delta}_{zz} &=&4G\left( \cos \varphi \left( t\right) \Delta
_{yz}+\sin \varphi \left( t\right) \Delta _{xz}\right)
-2T_1^{-1}\left( \Delta
_{zz}-n\right) , \\
\dot{\Delta}_{xy} &=&\omega _0\left( \Delta _{xx}-\Delta
_{yy}\right) -2G\cos \varphi \left( t\right) \Delta _{xz}-2G\sin
\varphi \left( t\right)
\Delta _{yz} \\
&&+q\left( \Delta _{xx}s_z-\Delta _{yy}s_z+\Delta _{xz}s_x-\Delta
_{yz}s_y\right)  \\
&&-2\left( 2T_2^{-1}-T_1^{-1}\right) \Delta _{xy}-4\left(
T_2^{-1}-T_1^{-1}\right) s_xs_y, \\
\dot{\Delta}_{xz} &=&-\omega _0\Delta _{yz}-q\Delta _{yz}s_z-q\Delta
_{zz}s_y+2G\cos \varphi \left( t\right) \Delta _{xy} \\
&&-2G\sin \varphi \left( t\right) \left[ \Delta _{zz}-\Delta
_{xx}\right]
-\left[ T_2^{-1}+T_1^{-1}\right] \Delta _{xz}, \\
\dot{\Delta}_{yz} &=&\omega _0\Delta _{xz}+q\left( \Delta
_{xz}s_z+\Delta
_{zz}s_x\right) +2G\cos \varphi \left( t\right) (\Delta _{yy} \\
&&-\Delta _{zz})+2G\sin \varphi \left( t\right) \Delta _{xy}-\left[
T_2^{-1}+T_1^{-1}\right] \Delta _{yz}.
\end{eqnarray*}

\begin{figure}[tbp]
\includegraphics[width=0.4\textwidth]{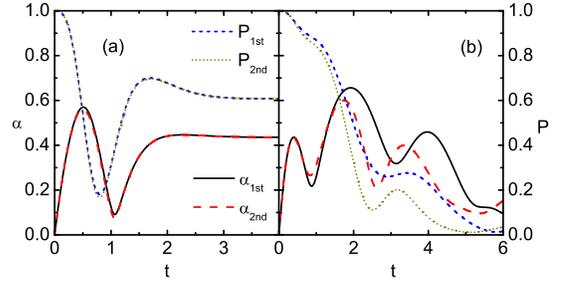}\newline
\caption{Comparison between the lowest and second order mean-field
approximation. q=0.001, $\omega_0$=0, $G_0$=1, $\Delta$=0, $f_a$
=0.5. (a) $\gamma$=2.0, $T_1^{-1}$=3.0; (b) $\gamma$=0.5,
$T_1^{-1}$=0.5.} \label{fig8}
\end{figure}
Fig. 8a and 8b display the coherence $\alpha$ and purity $P$
obtained based on both the lowest-order and second-order
approximation. For strong noise and dissipation case, as shown in
Fig. 8a, there is not obvious difference between the lowest-order
and second-order results. For weak noise and dissipation case,
although the second correction becomes more and more significant
with the time, the tendencies of the curves are similar [see Fig.
8b]. Therefore, we conclude that the lowest order approximation are
creditable qualitatively in the full parameter regime, especially in
the presence of the noise and dissipation.

\section{Summary}

In conclusion, we have investigated dynamics of the open
two-component BEC in the Bloch representation. The two components
correspond to two hyperfine states of the same specie of atom which
are coupled by the detuned Raman laser. We have calculated the
evolving Bloch vector, coherence and purity for both the closed
system and open system.

For the closed system, when transition frequency match with Raman
laser, the evolving Bloch vector form a closed orbital in the
surface of Bloch sphere. This mean that all physical quantities for
this system evolve periodically. As the nonlinear term becomes
strong the MQST exhibits. In the detuning case, the orbital shall
not be closed and wind into a basketlike path. In comparison with
the non-detuning case, although atoms still transit between two
states periodically, its transition period become longer. For the
system subject to noise the atoms always evolve to a balance state
that corresponds to the zero point in the Bloch space. In the
situation of strong nonlinear term and large detuning the orbital of
Bloch vector shall be irregular. For the system of dissipation,
along with atoms oscillate aperiodically between two states more and
more atoms escape from the condensates. The evolving Bloch vector
behave as an irregular orbital in the surface of Bloch sphere. When
the condensates subject to both noise and dissipation the system
always evolve to the same final state although the evolving path
might be different for different noise strength and dissipation
rate. It deserves to notice that the detuning between transition
frequency and laser frequency change obviously the dynamics of the
closed system, while the detuning effect is not displayed obviously
in the presence of noise and dissipation.

We also studied the coherence and purity for both closed and open
system. For the closed system the coherence always oscillates
periodically. As subject to noise the coherence shall decay to zero
during the period of evolution, while the particle loss induce no
obvious effect on the coherence. By tuning the noise strength and
dissipation rate the interesting effect of coherence enhancement
exhibits. The investigation of purity show that the system preserve
the maximum purity as long as it is not subject to noise even if the
dissipation exists. In spite that the noise shall decrease the
purity, it can be enhanced to a large value by tuning the noise
strength and dissipation.

We discussed the second-order correction of the master equations.
Comparison between the results from the lowest-order approximation
with those from the second-order one has shown that the former can
describe qualitatively the dynamics of both open and closed BECs and
with the enhancement of coupling with environment its quantitative
result shall be exact.

\begin{acknowledgments}
The authors acknowledge the National Natural Science Foundation of
China (Grant No. 11004007 and Grant No. 11074021) for financial
support. This work is also supported by the Fundamental Research
Funds for the Central Universities of China.
\end{acknowledgments}

\end{document}